\documentclass{article}

\usepackage[english]{babel}
\usepackage{abstract}
\usepackage{booktabs}
\usepackage{tabularx}
\usepackage[letterpaper,top=2cm,bottom=2cm,left=3cm,right=3cm,marginparwidth=1.75cm]{geometry}

\usepackage{amsmath}
\usepackage{amsfonts}
\usepackage{graphicx}
\usepackage[colorlinks=true, allcolors=blue]{hyperref}
\usepackage{parskip}
\usepackage{tcolorbox}

\title{Evaluating AI Recruitment Sourcing Tools by Human Preference}

\author{Vladimir Slaykovskiy, Maksim Zvegintsev, Yury Sakhonchyk, Hrachik Ajamian}

\begin{document}
\maketitle

\begin{abstract}
This study introduces a benchmarking methodology designed to evaluate the performance of AI-driven recruitment sourcing tools. We created and utilized a dataset to perform a comparative analysis of search results generated by leading AI-based solutions, LinkedIn Recruiter, and our proprietary system, Pearch.ai. Human experts assessed the relevance of the returned candidates, and an Elo rating system was applied to quantitatively measure each tool's comparative performance. Our findings indicate that AI-driven recruitment sourcing tools consistently outperform LinkedIn Recruiter in candidate relevance, with Pearch.ai achieving the highest performance scores. Furthermore, we found a strong alignment between AI-based evaluations and human judgments, highlighting the potential for advanced AI technologies—particularly our proposed solution—to substantially enhance talent acquisition effectiveness. Code and supporting data are publicly available at \href{https://github.com/vslaykovsky/ai-sourcing-benchmark}{https://github.com/vslaykovsky/ai-sourcing-benchmark}.

\end{abstract}

\section{Introduction}

For decades, talent acquisition professionals have relied on traditional search and filtering systems to identify suitable candidates. These legacy tools typically require users to possess a significant amount of skill and familiarity with search operations, often relying on Boolean logic, keyword lists, and numerous filtering parameters. The process can be time-consuming and iterative, as recruiters refine searches or experiment with different combinations of terms and filters in hopes of finding the right pool of candidates. Moreover, because these systems lack the capacity to interpret nuanced, context-rich queries, certain candidate attributes—such as having led a sales team of ten or more people—are more difficult to capture using standard filter-based approaches alone.

Recent developments in artificial intelligence, however, have paved the way for new sourcing tools that better understand and apply natural language processing techniques. With RAG systems \cite{lewis2020retrieval} that interpret the semantic content of job descriptions and candidate profiles, these AI-driven platforms often produce search results that are more relevant and more easily refined without requiring elaborate Boolean queries. Despite the excitement surrounding these novel technologies, the question remains: How much better do these AI-enabled systems actually perform compared to established market leaders? And can they truly overcome the limitations of traditional methods?

This study aims to quantify the performance differences by benchmarking leading AI-based sourcing tools against LinkedIn Recruiter—commonly regarded as the gold standard in the industry. By building a dataset of search queries, comparing search results side by side, and assessing these results through human expert evaluations and LLM-judge \cite{zheng2023judging}, this study employs an Elo rating system \cite{elo1966uscf} to measure the relative effectiveness of each platform. In doing so, we hope to demonstrate whether AI-driven tools can significantly improve hiring efficiency by reducing the time and effort required to find the most suitable candidates.
In particular, there are currently no comprehensive benchmarks or comparative analyses that assess these systems in a systematic way.

\section{Selection of Sourcing Tools}
 
For this study, we selected four sourcing tools based on their specialization as sourcing search engines. Specifically, the tools evaluated were \textbf{LinkedIn Recruiter}, \textbf{JuiceBox (PeopleGPT)}, \textbf{Exa.ai}, and our internally developed system, \textbf{Pearch.ai}.

\begin{itemize}
    \item \textbf{LinkedIn Recruiter}: Widely recognized as the leader of the industry, with over 1 billion total users.\\
    \url{https://news.linkedin.com/about-us#Statistics}
    
    \item \textbf{JuiceBox (PeopleGPT)}: One of the fastest-growing people search engines.\\
    \url{https://www.crunchbase.com/organization/peoplegpt}
    
    \item \textbf{Exa.ai}: An advanced search engine that has raised a total of \$22 million in funding.\\
    \url{https://www.crunchbase.com/organization/exa-1b30}
    
    \item \textbf{Pearch.ai}: Our proprietary people search engine, the result of two years of active R\&D.\\
    \url{https://pearch.ai}
\end{itemize}

We acknowledge that this selection is not exhaustive, but we believe it to be representative of the state of the industry at the time of this study. We intend to include additional tools in future iterations of this research.

\section{Selection of Search Queries}

To construct a robust benchmark, search queries were sourced from anonymized user traffic data on \texttt{Pearch.ai}. To ensure both diversity and uniqueness, we applied text embeddings \cite{reimers2019sentence} followed by $k$-means clustering to group semantically similar queries. Representative queries were then selected based on the centroids of each cluster. This approach reduced redundancy, specifically filtering out reformulated queries frequently generated by users of \texttt{Pearch.ai}.

To maintain fairness in evaluation, the selected queries were excluded from the training data of the \texttt{Pearch.ai} system. Furthermore, since the \texttt{Pearch.ai} index primarily contains U.S.-based profiles at the time of publication, we modified some queries by appending the constraint \emph{"Only show people from the U.S."} when geographic location was not explicitly stated by the user.

In total, we compiled 48 queries for this study.  

\subsection*{Example Search Queries}

\begin{itemize}
    \item \textit{A backend engineer from a top computer science program living in san francisco with 5 years experience working in startups developing performant and scalable backend systems and databases}
    \item \textit{founders who sold their companies or did an IPO. Only show people from the U.S.}
    \item \textit{Customer success managers with at least 3 years of experience who are located in the San Francisco Bay Area. I would like them to come from companies that are Series A, Series B, or Series C. They need to have worked in healthcare technology at a SAAS company.}
\end{itemize}

\section{Collection of Search Results}

To ensure consistent conditions and circumvent the lack of APIs in some of the tools, we used the web UI rather than APIs across all tested sourcing tools.

For LinkedIn Recruiter, we took advantage of its latest AI-assisted search feature\footnote{\url{https://business.linkedin.com/talent-solutions/ai-assisted-search-and-projects}}. In cases where initial searches returned no results, we clicked "Enhance this search based on past successful searches" up to three times to maximize the likelihood of getting any search results. This feature works by progressively relaxing the strict search criteria—broadening keyword matches, considering related skills and job titles, and loosening geographic or experience requirements—to capture a wider pool of candidate profiles.  For all other tools, search queries were executed as-is, collecting up to the top ten candidate profiles along with their corresponding rankings.

We reduced the outputs of all the systems down to LinkedIn profile links to ensure uniformity and simplicity of subsequent evaluations.

In total, we collected 1,735 search results. See Table~\ref{tab:results-summary} for more details.

\begin{table}[h!]
\centering
\begin{tabular}{|l|c|c|c|c|}
\hline
\textbf{Source} & \textbf{Queries} & \textbf{Total Results} & \textbf{Average per Query} & \textbf{Subscription Cost} \\
\hline
Exa.ai       & 48 & 465 & 9.7 & 800\$ \\
\hline
LinkedIn Recruiter     & 48 & 465 & 9.7 & 900\$\\
\hline
Pearch.ai    & 45 & 398 & 8.8 & 400\$ \\
\hline
JuiceBox    & 42 & 407 & 9.7 & 150\$ \\
\hline
\end{tabular}
\caption{Summary of search queries and result volumes per sourcing tool}
\label{tab:results-summary}
\end{table}

We observed that discrepancies between the number of total search results are explained by the following factors:
\begin{itemize}
    \item Some systems returned fewer than 10 search results for particular queries.
    \item Some LinkedIn links could not be opened and were therefore discarded.
\end{itemize}

\section{Comparing Search Results}

We used both a panel of human expert recruiters and \texttt{LLM-judge} \cite{zheng2023judging} to evaluate the search results. In either case, we asked an annotator (human or AI) to compare profiles at the same position in search results of the same query across different systems. 

For this, we generated a uniform random sample of 1,000 records consisting of the tuple: \texttt{(Query, Rank, Source1, DocID1, Source2, DocID2)} from the dataset of search results. See Table \ref{tab:dataset_sample} for a short sample of the resulting table. 

\begin{table}[h]
\centering
\begin{tabular}{|>{\raggedright\arraybackslash}p{4cm}|c|c|c|c|c|}
\hline
\textbf{Query} & \textbf{Rank} & \textbf{Source1} & \textbf{DocID1} & \textbf{Source2} & \textbf{DocID2} \\
\hline
python developer, russian speaking, 7+ years of experience. Only show people from the U.S. & 6 & JuiceBox & vegasq & Exa.AI & ana7pana\\
\hline
A perception or computer vision engineer with experience at a startup related to self-driving cars or drones based in San Francisco & 1 & LinkedIn & jayant1408 & JuiceBox & nghia-ho-6259866\\
\hline
\end{tabular}
\caption{Sample of the produced dataset.}
\label{tab:dataset_sample}
\end{table}

\subsection{Evaluation with LLM-judge}

We evaluated samples for the dataset above using the following ChatGPT prompt:

\begin{tcolorbox}[colback=gray!10, colframe=gray!80, title=Evaluation Prompt]
\begin{verbatim}
I'll give you a search query for a working professional and two candidate
profiles. Your task is to decide which profile is the better match.  
Make sure to consider all specific requirements in the query. 
Also, check if a candidate is overqualified or underqualified.
Output your decision as JSON: {
  'profile_idx': int # 1 or 2
  'summary': str # very short summary explaining the decision
}

Search query: {{query}}


Profile 1: {{profile1}}

Profile 2: {{profile2}}
\end{verbatim}
\end{tcolorbox}

Template variables \texttt{query}, \texttt{profile1}, \texttt{profile2} were dynamically replaced with actual query data and profile JSON representations using the following rules:

\begin{itemize}
  \item Queries were passed to evaluation engines as is, without any changes.
  \item Candidate profiles were structured as JSON documents sourced from LinkedIn profile pages. To enrich decision-making, these profiles were augmented with detailed company information extracted from Crunchbase \footnote{\url{https://crunchbase.com}}, including industry, company size, revenue, age, funding rounds, and other relevant information.
\end{itemize}

The OpenAI model (\texttt{o3-mini-2025-01-31}) produced the binary preference in the \texttt{profile\_idx} field with additional explanation in the \texttt{summary} field.

\subsection{Panel of Human Experts}

We selected a panel of eight expert recruiters through Upwork\footnote{\href{https://upwork.com}{https://upwork.com}}. The selection was based on the following criteria:
\begin{itemize}
    \item Location: Based in the United States
    \item Relevant experience: Titles related to recruiting, talent acquisition, sourcing, or human resources
    \item Proven success: Completed previous recruiting projects with a high success rate on Upwork
    \item Cost: Hourly rates between \$25 and \$50
\end{itemize}

The recruiters demonstrated a range of professional capabilities:
\begin{itemize}
    \item Hires per month ranged from 2--3 to as many as 20--25
    \item Simultaneous roles handled varied between 2 and 45
    \item Industry experience included healthcare, IT, finance, manufacturing, and more
    \item Candidate levels spanned from junior to executive positions
\end{itemize}

\begin{tcolorbox}[colback=gray!10, colframe=gray!80, title=Instruction given to each expert]
For each query, you will see two LinkedIn profiles. Read the query carefully and compare all its meaningful parts with the profiles. Sometimes the query includes extra details about a company. If needed, check company info on LinkedIn company pages, the company website, CrunchBase, or other sources you normally use. Choose the profile that best fits the query. In the ``Winner'' column, choose the winner (A or B candidate). In the ``Comment'' column, leave a short note explaining your choice.
\end{tcolorbox}

To ensure annotation quality, each recruiter was also given a test batch of queries. This batch was not labeled as a test in order to minimize performance inconsistencies between test and real assignments. Recruiter outputs were evaluated against a consensus-based Majority Vote to assess alignment quality. 

Test task statistics are shown in Table~\ref{tab:test_stats}. Note that recruiter names have been anonymized to protect their privacy. Based on performance, we selected the top 50\% of recruiters by majority alignment rate to form a group of high-quality annotators, referred to as ``Top Humans.'' This group includes: \texttt{Recruiter002}, \texttt{Recruiter008}, \texttt{Recruiter003}, \texttt{Recruiter005}.

\begin{table}[ht]
\centering
\begin{tabular}{lccc}
\hline
\textbf{Worker} & \textbf{Alignment with Majority Vote} & \textbf{Total Samples} & \textbf{Total Queries} \\
\hline
Recruiter002 & 0.74 & 50 & 32 \\
Recruiter008 & 0.74 & 50 & 32 \\
Recruiter003 & 0.70 & 50 & 32 \\
Recruiter005 & 0.69 & 48 & 31 \\
Recruiter007 & 0.69 & 49 & 31 \\
Recruiter001 & 0.60 & 50 & 32 \\
Recruiter006 & 0.58 & 50 & 32 \\
Recruiter004 & 0.53 & 49 & 31 \\
\hline
\end{tabular}
\caption{Recruiter Performance Metrics}
\label{tab:test_stats}
\end{table}

\begin{table}[ht]
\centering
\begin{tabular}{lcc}
\hline
\textbf{Worker} & \textbf{Total Labeled Samples} & \textbf{Total Queries} \\
\hline
Recruiter001 & 100 & 38 \\
Recruiter002 & 96  & 42 \\
Recruiter003 & 100 & 42 \\
Recruiter004 & 99  & 41 \\
Recruiter005 & 98  & 40 \\
Recruiter006 & 100 & 40 \\
Recruiter007 & 200 & 47 \\
Recruiter008 & 100 & 42 \\
\hline
\end{tabular}
\caption{Statistics of samples collected from expert recruiters}
\label{tab:sample_stats}
\end{table}

The total cost of human annotations: \$3,079.

\subsection{Correlation between Human Experts and LLM-judge}

We analyzed the alignment between the majority vote of human experts and the judgments of the \texttt{LLM-judge}. For each recruiter, we computed two alignment scores:

\begin{itemize}
    \item \textbf{Alignment with Majority Vote}: This score reflects how often a recruiter agreed with the majority of human annotators. For each sample, we first determined the majority decision across all human votes (i.e., which profile—A or B—was preferred). Then, for each individual recruiter, we computed the proportion of samples where their selection matched this majority label. Formally, if a recruiter annotated $N$ samples and matched the majority decision in $k$ of them, their alignment score is $k/N$.
    
    \item \textbf{Alignment with LLM-judge}: This score captures how often a recruiter's preference matched the decision made by the LLM-based evaluation model. Similar to the majority vote alignment, we calculated the proportion of recruiter judgments that were consistent with the \texttt{LLM-judge} output for the same query and profile pair.
\end{itemize}

\begin{figure}[h]
\centering
\includegraphics[width=0.7\textwidth]{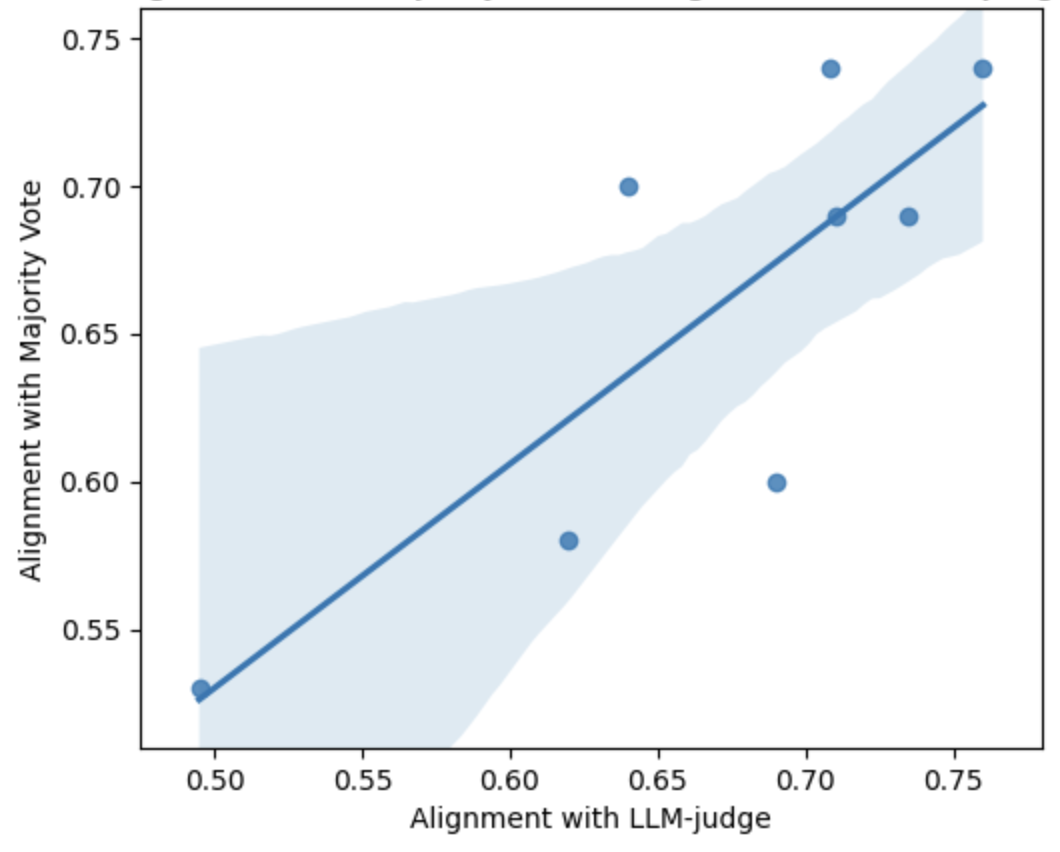}
\caption{Alignment with Majority Vote vs. Alignment with LLM-judge}
\label{fig:alignment_chart}
\end{figure}

Figure~\ref{fig:alignment_chart} illustrates the relationship between these two metrics for all eight recruiters. Each point represents one recruiter, with the x-axis showing their alignment with the \texttt{LLM-judge} and the y-axis showing their alignment with the majority vote.

A linear regression line fitted to the data reveals a strong positive correlation, suggesting that recruiters who more often agreed with the \texttt{LLM-judge} also tended to align better with the overall human consensus.

The shaded region in the chart denotes the 95\% confidence interval around the regression line, demonstrating the reliability of the observed trend. The Pearson correlation coefficient is 0.82 with a p-value of 0.0137, indicating a strong and statistically significant relationship. These results strongly support the validity of \texttt{LLM-judge} as an automated evaluation tool, capable of producing decisions that closely align with those of expert human recruiters.

\section{Ranking Sourcing Tools}

Our data comprises pairwise comparisons—yet how can we leverage these comparisons to infer a global ranking over all $M$ sourcing tools? This challenge is well-explored in the learning-to-rank literature \cite{liu2009learning}, and we provide our approach here. Let $\mathcal{A} = \{(m, m') : m < m', \; m, m' \in [M]\}$ represent the set of all possible tool pairs.

Following \cite{chiang2024chatbot}, we adopt a sequential framework in which, at each time step $t \in \mathbb{N}$, a pair of tools $A_t \in \mathcal{A}$ is presented to an evaluator (either human or AI), and we record their preference as $H_t \in \{0,1\}$. For instance, if $A_t = (1, 2)$ and $H_t = 1$, this means the evaluator preferred the output from tool 2 over that from tool 1.

\subsection{Win-rate Matrix Estimation}

A natural way to represent preference data is through the construction of a win-rate matrix. For any pair of sourcing tools $a = (a_1, a_2) \in \mathcal{A}$, we define the win-rate $\theta^*(a)$ as the probability that the evaluator prefers tool $a_2$ over $a_1$:

\[
\theta^*(a) = \mathbb{E}[H_t \mid A_t = a].
\]

Here, $H_t \in \{0, 1\}$ denotes the evaluator's response at time $t$, with $H_t = 1$ indicating a preference for the second tool in the pair.

We estimate the win-rate matrix using importance-weighted averages. For each time step $t$, define:

\[
X_t(a) = \frac{1}{P_t(a)} H_t \cdot \mathbb{1}\{A_t = a\},
\]

where $P_t(a)$ denotes the probability of sampling pair $a$ at time $t$. The unbiased estimator for the win-rate vector $\theta^*$ is then:

\[
\hat{\theta}_T = \frac{1}{T} \sum_{t=1}^{T} X_t. \tag{4}
\]

This estimator converges to the true win-rate as the number of comparisons $T$ increases, assuming sufficient sampling across all pairs.

To derive a global ranking from the win-rate matrix, we compute the average win probability for each tool by averaging over its row entries. This yields a score that reflects how frequently a tool wins when compared to others:

\[
\text{Score}(i) = \frac{1}{M - 1} \sum_{\substack{j=1 \\ j \neq i}}^{M} \hat{W}_{i,j},
\]

where $\hat{W}_{i,j} = \hat{\theta}_T((i, j))$ if $i < j$, and $1 - \hat{\theta}_T((j, i))$ if $i > j$.

\begin{figure}
\centering
\includegraphics[width=1\linewidth]{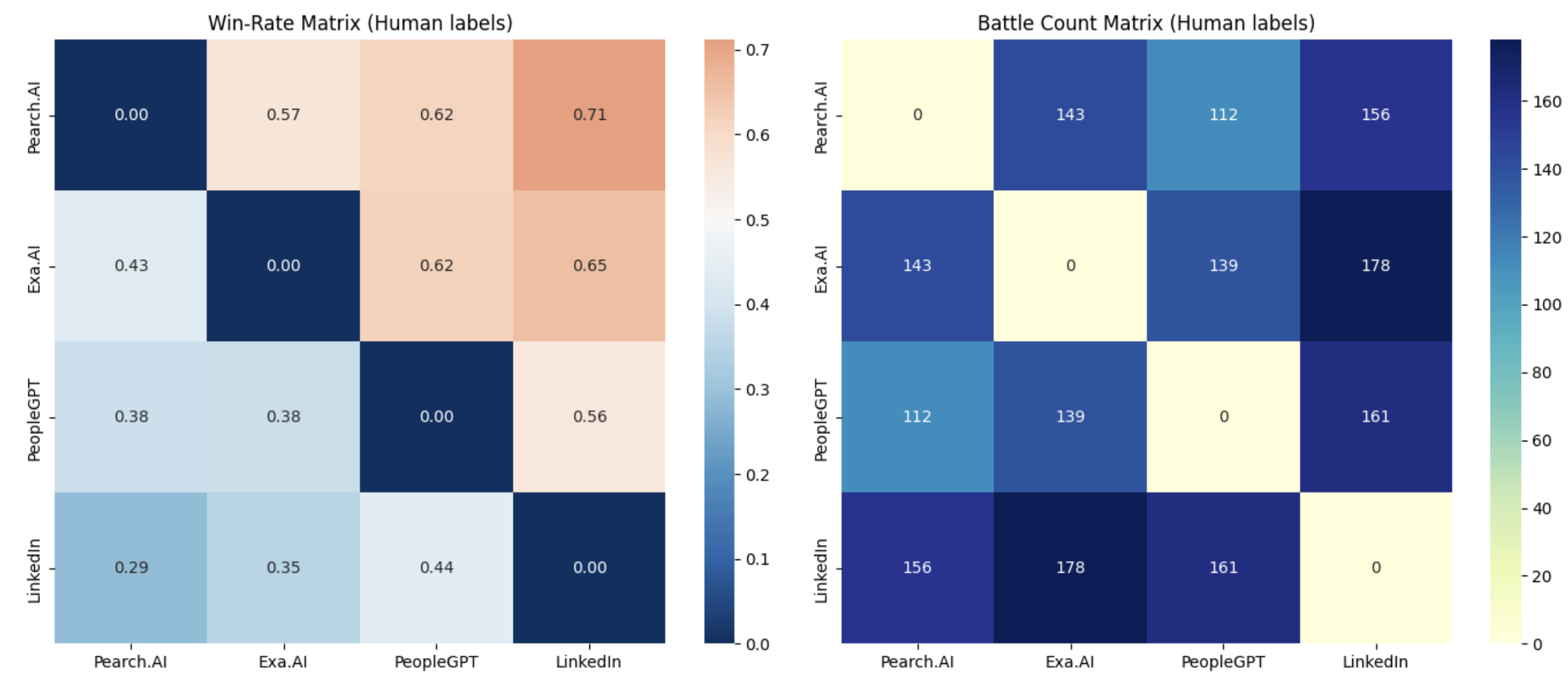}
\caption{\label{fig:win-rate-human}Win-rate Matrix based on Human Preference Data}
\end{figure}

\begin{figure}
\centering
\includegraphics[width=1\linewidth]{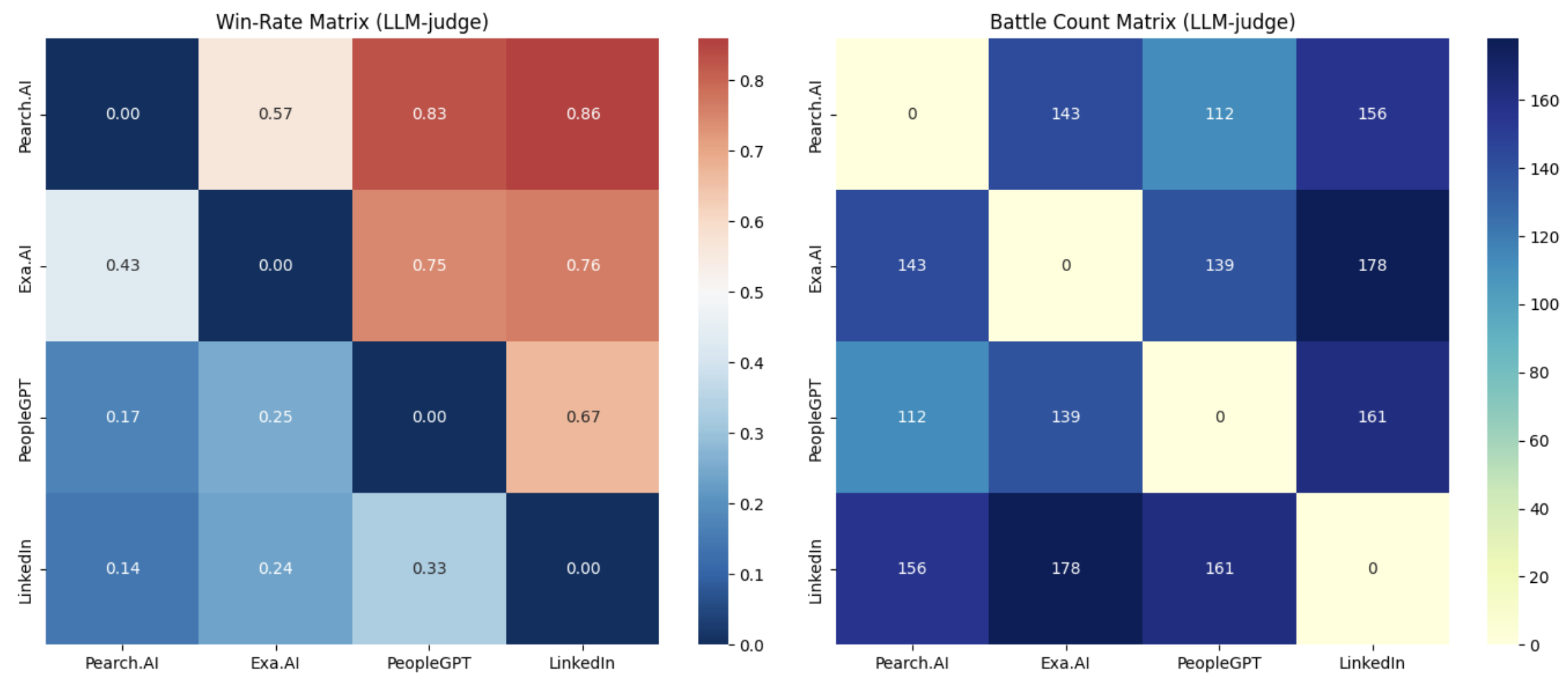}
\caption{\label{fig:win-rate-llm}Win-rate Matrix based on Preference Data of LLM-Judge}
\end{figure}

Figures~\ref{fig:win-rate-human} and~\ref{fig:win-rate-llm} present the empirical win-rate matrices computed from human expert annotations and \texttt{LLM-judge} preferences, respectively. Both matrices produce the same tool ordering, offering compelling evidence that state-of-the-art LLMs exhibit preference patterns highly consistent with expert human judgments in the context of candidate relevance.

It is important to note, however, that win-rate matrices do not necessarily produce transitive preference structures. That is, it is possible for Tool A to outperform Tool B, Tool B to outperform Tool C, and Tool C to outperform Tool A. Such cycles introduce ambiguity into global ranking efforts.

To address this, we turn to the Elo rating system, which produces a consistent, transitive ranking that best fits the observed pairwise preferences.

\begin{figure}
\centering
\includegraphics[width=0.5\linewidth]{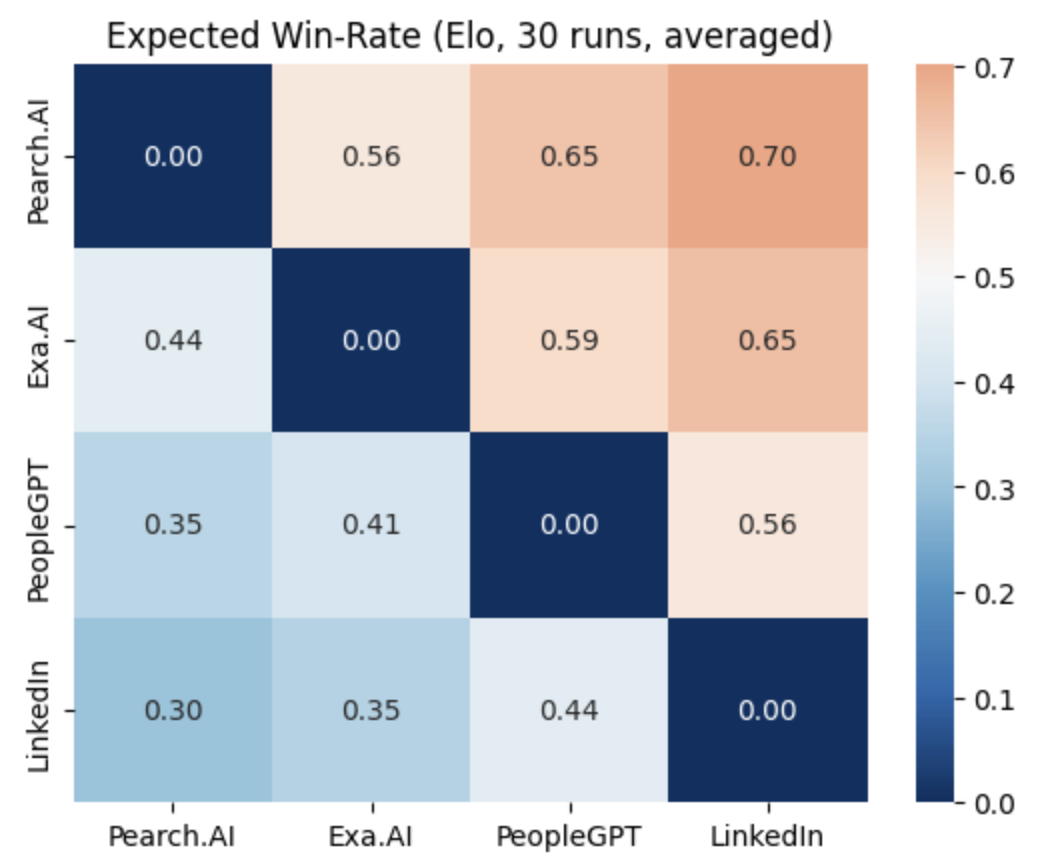}
\caption{\label{fig:win-rate-elo}Expected Win-rate Matrix based on Elo ratings}
\end{figure}

\subsection{Elo Rating System}

The Elo rating system~\cite{elo1966uscf}, originally developed for ranking chess players, has been widely adopted in machine learning research, including recent applications for evaluating Large Language Models (LLMs)~\cite{zheng2023judging}. In this study, we apply the Elo framework to derive relative performance scores for each sourcing tool based on the outcomes of pairwise comparisons.

We treat each comparison—whether judged by humans or an LLM—as a match between two sourcing tools. Let $R_A$ and $R_B$ denote the Elo ratings of tools $A$ and $B$, respectively. The expected probability that $A$ beats $B$ is given by the logistic function:

\[
\mathbb{E}[A \succ B] = \frac{1}{1 + 10^{(R_B - R_A)/400}}.
\]

After observing an outcome $S_A \in \{0, 1\}$—where $1$ indicates $A$ was preferred and $0$ otherwise—the Elo rating is updated as follows:

\[
R_A \leftarrow R_A + K (S_A - \mathbb{E}[A \succ B]),
\]
\[
R_B \leftarrow R_B + K ((1 - S_A) - \mathbb{E}[B \succ A]),
\]

where $K$ is a scaling factor that controls the sensitivity of the rating updates (we use $K = 40$ in our implementation).

To ensure robustness, we repeated the Elo calculation 30 times using randomized orderings of the comparison data, averaging results to produce stable estimates. Additionally, we computed the expected win-rate matrix derived from the final Elo scores (Figure~\ref{fig:win-rate-elo}):

\[
\hat{W}_{i,j} = \frac{1}{1 + 10^{(R_j - R_i)/400}},
\]

for each pair of tools $i$ and $j$. This matrix closely mirrors the empirical win-rate matrices in Figures~\ref{fig:win-rate-human} and~\ref{fig:win-rate-llm}, suggesting that the Elo model fits the underlying data well.

We evaluated Elo ratings using three different supervision sources:

\begin{itemize}
    \item \textbf{LLM-judge labels}: Evaluations produced by an automated LLM (o3-mini-2025-01-31).
    \item \textbf{Human labels}: Annotations from a general pool of expert recruiters.
    \item \textbf{Top human labels}: A filtered set of the most reliable recruiters based on majority-vote alignment.
\end{itemize}

\begin{figure}
\centering
\includegraphics[width=1\linewidth]{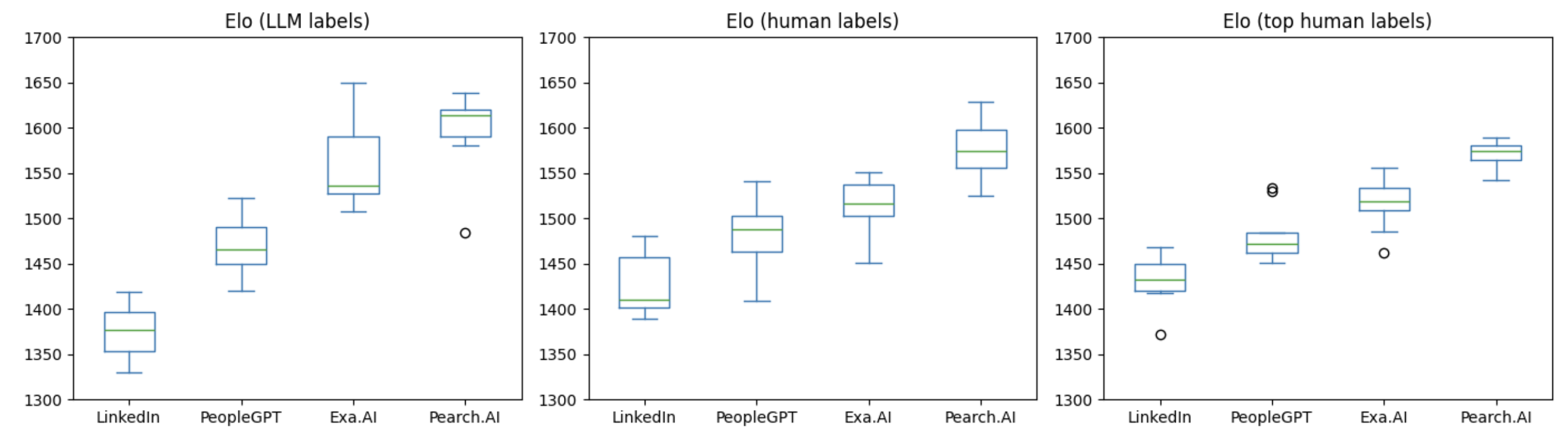}
\caption{\label{fig:elo}Elo diagrams}
\end{figure}

The final Elo scores across all three sources (Figure~\ref{fig:elo}) yield consistent rankings, with AI-native tools outperforming LinkedIn Recruiter in all settings. Notably, \texttt{Pearch.ai} achieves the highest Elo score, underscoring the effectiveness of domain-specialized AI in sourcing tasks.

\section{Conclusion}

Our study provides strong empirical evidence that AI-based sourcing tools outperform traditional recruitment platforms in terms of candidate relevance and alignment with user queries. Among the evaluated systems, \texttt{Pearch.ai} achieved the highest performance scores, suggesting that carefully designed, domain-specialized AI systems can significantly exceed the capabilities of established industry solutions such as LinkedIn Recruiter.

A notable contribution of this work is the demonstration of \texttt{LLM-judge} as a reliable evaluator for sourcing tool comparisons. Its evaluations show strong alignment with those of expert human annotators, supporting the idea that modern large language models can serve not only as components of search systems but also as effective automated judges in benchmarking pipelines. This automation can substantially reduce the cost, time, and subjectivity associated with human annotation, enabling more scalable and continuous evaluations.

While our results are promising, they also highlight several important directions for future research. First, our evaluation primarily focused on precision—how well the top results matched the given query. Future studies should incorporate recall, measuring how comprehensively systems identify relevant candidates across the full result set.

Second, although we included subscription pricing as a rough proxy for system cost, a more rigorous evaluation of compute efficiency—such as the dollar cost per relevant candidate—is essential. As AI systems become increasingly embedded in operational workflows, understanding these cost-performance tradeoffs will be critical.

Lastly, this benchmark involved a limited number of tools and search queries. Future work should expand the range of sourcing systems evaluated, increase dataset size, and include a broader diversity of job roles and query formulations. We believe that building larger, standardized benchmarks—combined with automated evaluators like \texttt{LLM-judge}—will play a pivotal role in accelerating progress in AI-assisted talent acquisition.

In summary, AI sourcing tools are already capable of outperforming traditional platforms, and with robust evaluation frameworks, we are well-positioned to optimize and scale the next generation of recruitment technologies.

\section*{Disclosure}

The authors are affiliated with Pearch.ai, one of the sourcing tools evaluated in this study. Although every effort has been made to ensure objectivity, including the exclusion of training data for selected queries and the use of third-party expert annotations, all readers should be aware of this potential conflict of interest. All evaluation procedures, data sources, and code have been made publicly available to support transparency and reproducibility.

\bibliographystyle{alpha}
\bibliography{bibliography}

\end{document}